\documentclass[prl,twocolumn,superscriptaddress]{revtex4-2}

\usepackage{booktabs} 
\usepackage{array}       
\usepackage{cellspace}   
\usepackage{graphicx}
\usepackage{float}
\usepackage{amsmath}
\usepackage{amssymb}
\usepackage{amsthm}
\usepackage{amsfonts}
\usepackage{listings}
\usepackage{tabularx}
\usepackage{makecell}
\usepackage{soul, color,xcolor}
\usepackage{enumerate}
\usepackage{latexsym}
\usepackage{color}
\usepackage{bm}
\usepackage{hyperref}
\hypersetup{
 pdfnewwindow=true, colorlinks=true,
 linkcolor=blue, anchorcolor=blue,
 citecolor=blue, filecolor=blue,
 menucolor=blue, urlcolor=blue}

\usepackage{psfrag}

\usepackage{bm}
\usepackage{graphicx}
\usepackage{subfigure}

\soulregister{\cite}7
\soulregister{\ref}7
\RequirePackage[normalem]{ulem} 
\RequirePackage{color}\definecolor{RED}{rgb}{1,0,0}\definecolor{BLUE}{rgb}{0,0,1} 

\date{April 2025}

\begin{document}

\tolerance 10000

\newcommand{\vk}{{\bf k}}

\draft

\title{SDW driven ``magnetic breakdown" in a d-wave altermagnet KV\textsubscript{2}Se\textsubscript{2}O}


\author{Xu Yan}
\affiliation{College of Physics and Hebei Advanced Thin Films Laboratory, Hebei Normal University, Shijiazhuang, Hebei 050024, China}
\affiliation{Beijing National Laboratory for Condensed Matter Physics,
and Institute of Physics, Chinese Academy of Sciences, Beijing 100190, China}

\author{Ziyin Song}
\affiliation{Beijing National Laboratory for Condensed Matter Physics,
and Institute of Physics, Chinese Academy of Sciences, Beijing 100190, China}
\affiliation{University of Chinese Academy of Sciences, Beijing 100049, China}

\author{Juntao Song}
\affiliation{College of Physics and Hebei Advanced Thin Films Laboratory, Hebei Normal University, Shijiazhuang, Hebei 050024, China}

\author{Zhong Fang}
\affiliation{Beijing National Laboratory for Condensed Matter Physics,
and Institute of Physics, Chinese Academy of Sciences, Beijing 100190, China}
\affiliation{University of Chinese Academy of Sciences, Beijing 100049, China}

\author{Hongming Weng}
\affiliation{Beijing National Laboratory for Condensed Matter Physics,
and Institute of Physics, Chinese Academy of Sciences, Beijing 100190, China}
\affiliation{University of Chinese Academy of Sciences, Beijing 100049, China}

\author{Quansheng Wu}
\email{quansheng.wu@iphy.ac.cn}
\affiliation{Beijing National Laboratory for Condensed Matter Physics,
and Institute of Physics, Chinese Academy of Sciences, Beijing 100190, China}
\affiliation{University of Chinese Academy of Sciences, Beijing 100049, China}

\begin{abstract}
Altermagnets, combining zero net magnetization with intrinsic spin splitting, demonstrate unique quantum phenomena crucial for spintronic applications. KV\textsubscript{2}Se\textsubscript{2}O is proven to be a d-wave altermagnet with phase transition from a checkerboard-type (C-type) antiferromagnetic (AFM) state to a spin density wave (SDW) state as the temperature decreases. After phase transition, the apparent paradox emerges where angle-resolved photoemission spectroscopy (ARPES) reveals negligible Fermi surface modifications, while physical property measurement system (PPMS) measurements uncover substantial changes in transport properties. Our study explores the microscopic mechanisms governing phase-dependent transport properties of KV\textsubscript{2}Se\textsubscript{2}O base on first-principles calculations. The spin canting driven by periodic spin modulation in the SDW phase reduces the magnetic symmetry of KV\textsubscript{2}Se\textsubscript{2}O. The resultant band degeneracy lifting and Fermi surface reconstruction induce the ``magnetic breakdown" phenomenon, which alters carrier trajectories, modifies carrier concentration, strengthens electron-hole compensation, and ultimately accounts for the contrasting magnetic-field-dependent Hall resistivity relative to the C-type AFM state. Our work proposes an innovative method for identifying the electronic structure evolution across phase transitions from transport signatures, providing a novel paradigm for altermagnets research.
\end{abstract}

\maketitle

\textit{Introduction}—Altermagnets, which have zero net magnetic moment and intrinsic non-relativistic spin band splitting owing to the unique symmetry between spin opposite sublattices (preserving rotational/mirror symmetry, while breaking both inversion and translational symmetries), exhibit various spin-dependent effects previously considered exclusive to ferromagnets, including the anomalous Hall effect\cite{vsmejkal2020crystal}, spin Nernst effect\cite{cheng2016spin}, giant magnetoresistance (MR)\cite{vsmejkal2022giant,shao2021spin}, spin-splitting torque\cite{bai2022observation}, and unconventional thermal responses\cite{vsmejkal2023chiral,cui2023efficient,zhou2024crystal}, positioning altermagnets as promising materials for next-generation technologies ranging from ultra-dense spintronic memory to quantum computing devices. The momentum dependent spin band splitting has been observed in diverse materials\cite{bai2024altermagnetism,song2025altermagnets}, such as \textit{d}-wave altermagnets Mn\textsubscript{5}Si\textsubscript{3}\cite{reichlova2024observation,rial2024altermagnetic}, MnTe\cite{krempasky2024altermagnetic,lee2024broken}, MnTe\textsubscript{2}\cite{zhu2024observation}, Rb\textsubscript{1-$\delta$}V\textsubscript{2}Te\textsubscript{2}O\cite{zhang2025crystal}, and \textit{g}-wave altermagnets CrSb\cite{reimers2024direct,zhou2025manipulation}, RuO\textsubscript{2}\cite{fedchenko2024observation,karube2022observation,bai2022observation,zhu2019anomalous,lin2024observation,feng2022anomalous}. Additionally, numerous theoretically proposed candidates remain to be experimentally verified\cite{chen2024enumeration,chen2025unconventional,jiang2024enumeration,xiao2024spin,ma2021multifunctional,liu2024twisted,gao2025ai,che2024realizing,jin2024anomalous,zhu2023multipiezo}.

Recently, the \textit{d}-wave altermagnetism is ascertained by spin- and angle-resolved photoemission spectroscopy (SARPES) in the tetragonal KV\textsubscript{2}Se\textsubscript{2}O with checkerboard-type (C-type) antiferromagentic (AFM) order [Figs. S1(a) and S1(b) in Supplemental Material \cite{supp}], originating from its [\textit{C}\textsubscript{2}\textbar\textbar \textit{C}\textsubscript{4\textit{z}}] spin space symmetry\cite{jiang2025metallic}. As the temperature decreases, a metal-insulator transition occurs at 105 K, followed by a resurgence of metallic behavior at 100 K, resembling traditional spin density wave (SDW) transitions\cite{bai2024absence}. Despite the emergence of a small energy gap below the Fermi surface is observed via ARPES after the SDW phase transition, the transport-relevant Fermi surface shapes exhibits no detectable changes in measurements\cite{jiang2025metallic}. On the other hand, the temperature-dependent transport characteristics under magnetic field, particularly the Hall resistivity ($\rho_{yx}$), demonstrate a pronounced enhancement at the SDW critical temperature [Fig. 1(d)]\cite{bai2024absence}. The robust Fermi surface topology alongside starkly divergent temperature-driven transport behavior of KV\textsubscript{2}Se\textsubscript{2}O presents an intriguing paradox that has yet to be fully resolved.

\begin{figure*}[htbp]
    \centering
    \includegraphics[width=1\linewidth]{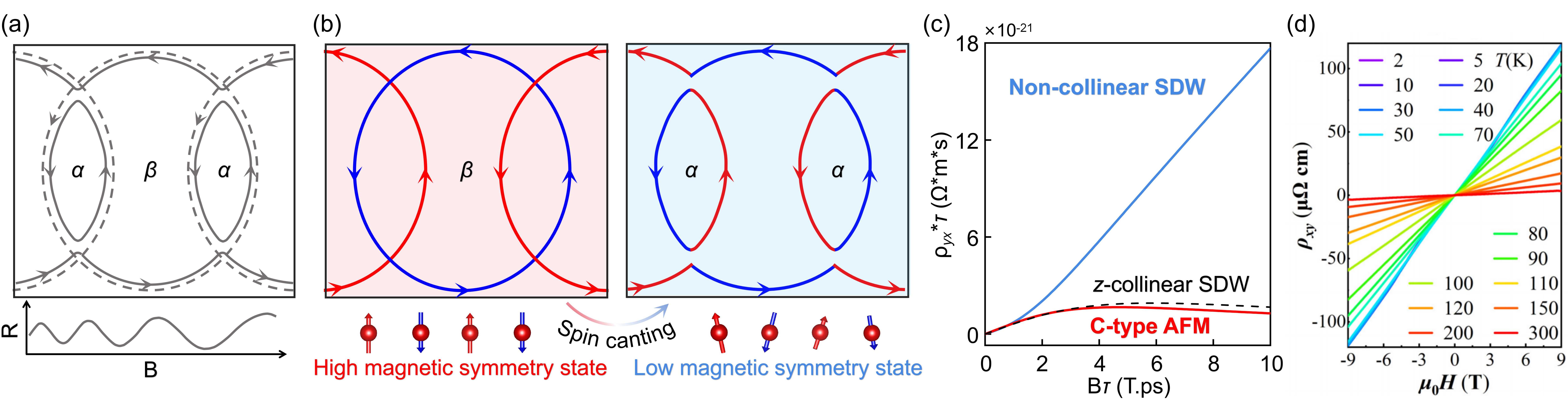}
    \caption{(a) Model Fermi surface of the quantum oscillation induced ``magnetic breakdown" phenomenon, in which the carriers motion along $\alpha$ and $\beta$ orbital under weak and strong magnetic field, respectively. (b) Model Fermi surface of the spin canting induced ``magnetic breakdown" phenomenon, the red and blue solid lines represent the spin-up and spin-down orbitals, respectively. (c) $\rho_{yx}$ of the KV\textsubscript{2}Se\textsubscript{2}O with C-type AFM, \textit{z}-collinear SDW, and non-collinear SDW spin alignment. (d) The measured temperature-dependent $\rho_{yx}$ of KV\textsubscript{2}Se\textsubscript{2}O\cite{bai2024absence}, copyright 2024, American Physical Society.}
    \label{fig:1}
   
\end{figure*}

Previous studies in various systems have demonstrated that the Fermi surface topology and the carrier motion trajectory play a crucial role in magnetotransport properties\cite{zhang2019magnetoresistance,ong1991geometric}. For instance, the compensation and open-orbit mechanisms in copper with single Fermi surface geometry contribute to magnetotransport resulting in an intricate angular MR diagram\cite{klauder1960proceedings}. The complex compensation between multi-valley Fermi surface gives rise to a distinct MR anisotropy pattern in the case of bismuth\cite{collaudin2015angle}. These findings suggest that even minor variations in electronic dispersion or band degeneracy—potentially undetected by experimental ARPES—might govern the emergent transport dichotomy in materials\cite{BIBERACHER2005360,ZhiXiaLi}. Therefore, unraveling the previously measured discrepancy of electronic and transport properties in KV\textsubscript{2}Se\textsubscript{2}O system demands a holistic exploration bridging spin alignment, electronic structure, and microscopic carrier dynamics.

In this work, we performed DFT calculations to confirm the electronic band structures and Fermi surface of C-type AFM and SDW KV\textsubscript{2}Se\textsubscript{2}O. Based on magnetic space group symmetry analysis, we attribute the phase-dependent transport dichotomy to spin-canting-induced symmetry breaking, which triggers a ``magnetic breakdown" in the SDW phase. This ``magnetic breakdown" phenomenon, illustrated in Fig. 1(b), is fundamentally distinct from the conventional ``magnetic breakdown" driven by quantum-oscillation-induced carriers motion change under strong magnetic fields, as previously proposed [Fig. 1(a)]. Boltzmann transport simulations under relaxation-time approximation yield $\rho_{yx}$ quantitatively consistent with experimental data. Combined with real-space electron trajectory visualization, these results confirm the decisive role of Fermi surface reconstruction via ``magnetic breakdown" in modifying transport behavior. Furthermore, the observed anisotropic MR in SDW KV\textsubscript{2}Se\textsubscript{2}O provides additional validation of our proposed mechanism.

\begin{figure*}[htbp]
    \centering
    \includegraphics[width=1.0\linewidth]{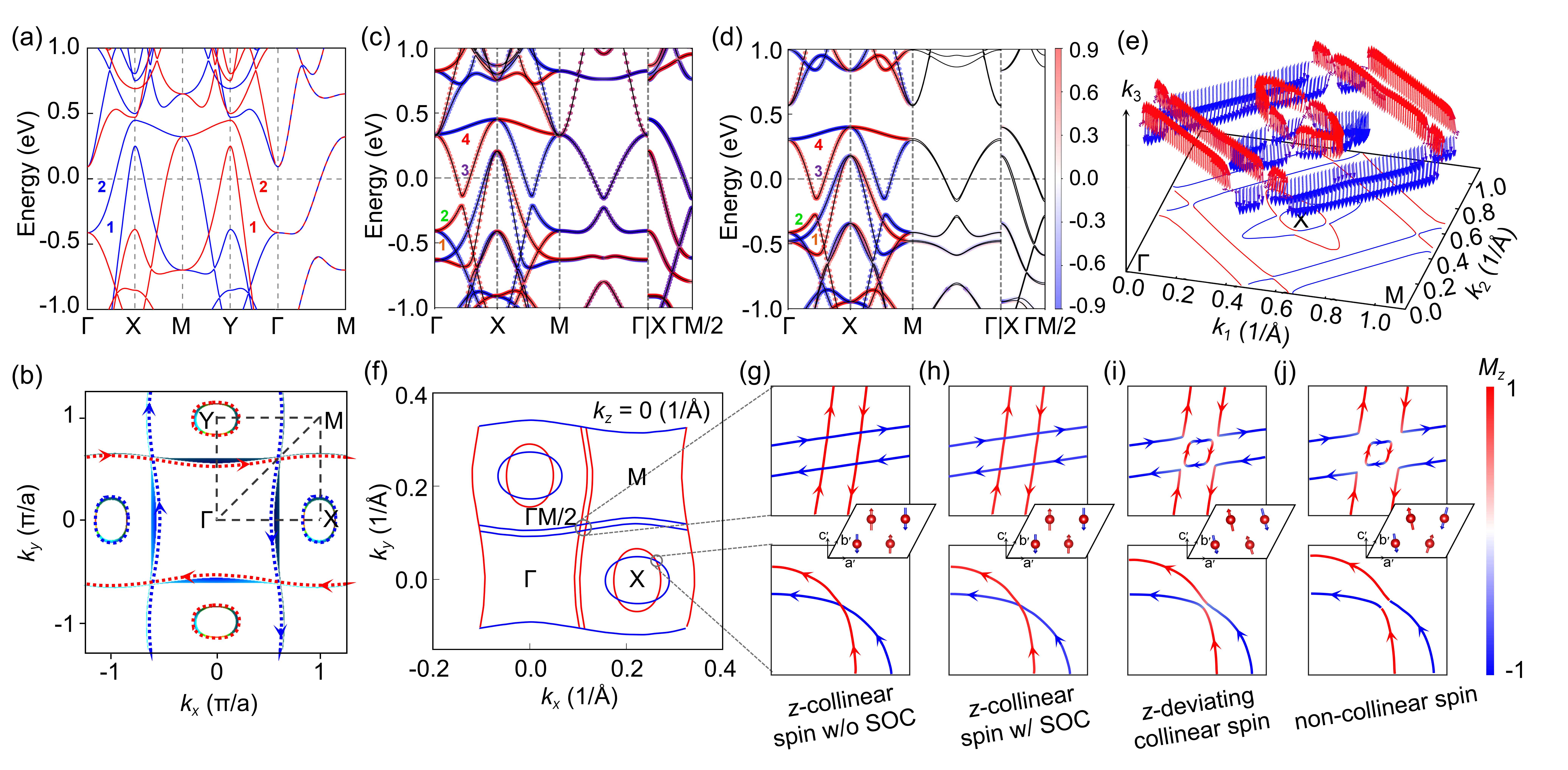}
    \caption{(a) Band structure and (b) Fermi surface of the C-type AFM KV\textsubscript{2}Se\textsubscript{2}O. The colored numbers resprent the band indices. Band structure of the SDW KV\textsubscript{2}Se\textsubscript{2}O with (c) \textit{z}-collinear and (d) \textit{z}-\textit{x}-non-collinear spin alignment including SOC interaction. (e) Spin textures at the Fermi level of non-collinear SDW KV\textsubscript{2}Se\textsubscript{2}O. (f) Fermi surface of the SDW KV\textsubscript{2}Se\textsubscript{2}O. Schematic illustration of the trajectory of the charge carriers in momentum space of the \textit{z}-collinear SDW KV\textsubscript{2}Se\textsubscript{2}O (g) without and (h) with considering SOC, SDW KV\textsubscript{2}Se\textsubscript{2}O (i) with \textit{z}-\textit{x}-collinear, and (j) \textit{z}-\textit{x}-non-collinear spin alignment, respectively.}
    \label{fig:2}
 \end{figure*}

\textit{Electronic properties}—The calculated band structure of the C-type AFM KV\textsubscript{2}Se\textsubscript{2}O reveals a pronounced spin splitting along the $\Gamma$-X-M and $\Gamma$-Y-M paths, and spin degeneracy along the $\Gamma$-M path, owing to the unique structural symmetry and spin ordering ([\textit{C}\textsubscript{2}\textbar\textbar \textit{C}\textsubscript{4\textit{z}}] spin group symmetry) [Fig. 2(a)], consistent with previous research\cite{jiang2025metallic}. The spin-up/down bands crossing the Fermi level along the $\Gamma$-X/Y and X/Y-M paths (donated as band-1) form small spin-polarized elliptical cylindrical Fermi surfaces, while those along the $\Gamma$-X/Y and $\Gamma$-M paths (donated as band-2) generates huge quasi-2D Fermi surfaces [Fig. 2(b)]. Surprisingly, strong anisotropy of the spin valley and Fermi surface implies substantial non-collinear spin current generation\cite{ma2021multifunctional}. 

For KV\textsubscript{2}Se\textsubscript{2}O below 100 K [Fig. S1(c)], the SDW phase transition introduces an energy gap approximately 0.15 eV below the Fermi level [Fig. S2(a)]. The persevered spin bands degeneration along the $\Gamma$-M and X-$\Gamma$M/2 paths can be attributed to the preservation of [\textit{C}\textsubscript{2}\textbar\textbar\textit{M}\textsubscript{1-10}] symmetry (Table S1). The [\textit{C}\textsubscript{2}\textbar\textbar \textit{C}\textsubscript{4\textit{z}}] symmetry breaking from periodic spin modulation induces intersecting Fermi surfaces, which have the same topology as that of the C-type AFM state [Figs. 2(b) and 2(f)]. Furthermore, the effect of SOC interaction on the spin degeneration in the SDW state with collinear spin alignment along \textit{z} direction (\textit{z}-collinear SDW) is criticized based on the magnetic space group theory analysis\cite{bradley1968magnetic}. The double degeneracies along the $\Gamma$-M and X-$\Gamma$M/2 paths under SOC conditions are fundamentally protected by [\textit{C}\textsubscript{2\textit{x}}\textbar \textit{t}\textsubscript{1/2}], [\textit{M}\textsubscript{001}\textbar \textit{t}\textsubscript{1/2}], [\textit{M}\textsubscript{010}], and [\textit{C}\textsubscript{2\textit{y}}], [\textit{M}\textsubscript{001}\textbar \textit{t}\textsubscript{1/2}], [\textit{M}\textsubscript{100}\textbar \textit{t}\textsubscript{1/2}] symmetry operations, respectively. These degenerate band features are also explicitly demonstrated in our first-principles calculations [Fig. 2(c)]. In particular, the \textit{z}-projected magnetic moment projections along these symmetry protected paths are approximately +1 and -1, respectively, representing strong spin polarization, similar to that without considering SOC. Therefore, the SOC interaction shows a negligible influence on the Fermi surface of \textit{z}-collinear SDW KV\textsubscript{2}Se\textsubscript{2}O [Figs. 2(g) and 2(h)]. 

\textit{Symmetry breaking induced ``magnetic breakdown"}— In order to explore the effect of periodic spin modulation on transport properties, we compare the magnetic field dependent $\rho_{yx}$ of the C-type AFM and \textit{z}-collinear SDW KV\textsubscript{2}Se\textsubscript{2}O [Fig. 1(c)]. The obvious deviation from experimental observations\cite{bai2024absence} indicates that the significant changes in magnetoelectric transport properties following the SDW phase transition cannot be solely attributed to the period magnetization variations of V atoms. On the other hand, the motion trajectory of carriers, which directly governed by Fermi surface topology, serves as a pivotal determinant of transport properties. If a small momentum gap opens at the spin-degenerate positions on the Fermi surface [Figs. 1(a) and 1(b)], the carrier trajectories will be entirely altered. From the perspective of group-theoretical analysis, this gap opening undoubtedly requires breaking the magnetic symmetry, lifting the spin band degeneracy along the $\Gamma$-M and X-$\Gamma$M/2 paths. 

Bearing this in mind, we construct SDW phases with reduced magnetic symmetry by rotating spin orientations, including collinear/non-collinear spin alignment from \textit{z} axis tilting to \textit{x} axis and \textit{xy} plane, respectively (donated as \textit{z}-\textit{x}-collinear/non-collinear and \textit{z}-\textit{xy}-collinear/non-collinear), to systematically investigate how symmetry breaking governs transport properties in the SDW KV\textsubscript{2}Se\textsubscript{2}O. Concretely, when the SDW KV\textsubscript{2}Se\textsubscript{2}O have \textit{z}-\textit{x}-collinear, \textit{z}-\textit{xy}-collinear, \textit{z}-\textit{x}-non-collinear, and \textit{z}-\textit{xy}-non-collinear spin alignment, the magnetic space group symmetry degrades from \textit{Pmma}.1 in \textit{z}-collinear SDW state to \textit{P}2\textsubscript{1}/\textit{m}.1, \textit{P}-1.1, \textit{P}2/\textit{c}.1, and \textit{P}-1.1, respectively. Such symmetry breaking remove all protective constraints along the $\Gamma$-M and X-$\Gamma$M/2 paths, resulting in complete splitting spin bands [Figs. 2(c), 2(d) and S2(b)-(d)]. The progressively broken magnetic symmetry in the SDW KV\textsubscript{2}Se\textsubscript{2}O with \textit{xy}-canting spin alignment results in enhanced band splitting than that in the \textit{x}-canting SDW states. Consequently, the emergence of momentum gaps at original Fermi surface crossing points drives a morphological reconstruction of the Fermi surface, manifesting as distinct geometrical configurations including near-circular profiles (band 1), petal-shaped contours (band 2), convex/concave quadrilaterals (band 3), and miniature parallelogram (band 4) shapes [Figs. 2(f), 2(i) and 2(j)]. Moreover, the Fermi surface reconstruction provides an effective channel for spin flipping. The spin of electrons quickly flips along the $\Gamma$-M and X-$\Gamma$M/2 paths, which is termed as ``magnetic breakdown" phenomenon here and can be testified by the spin textures at the Fermi level [Fig. 2(e)]. Therefore, the Fermi surface reconstruction related ``magnetic breakdown" phenomenon leads to disparate carrier motion trajectory in the SDW KV\textsubscript{2}Se\textsubscript{2}O compare to that in the C-type AFM state.

\begin{figure}[htbp]
    \centering
    \includegraphics[width=1\linewidth]{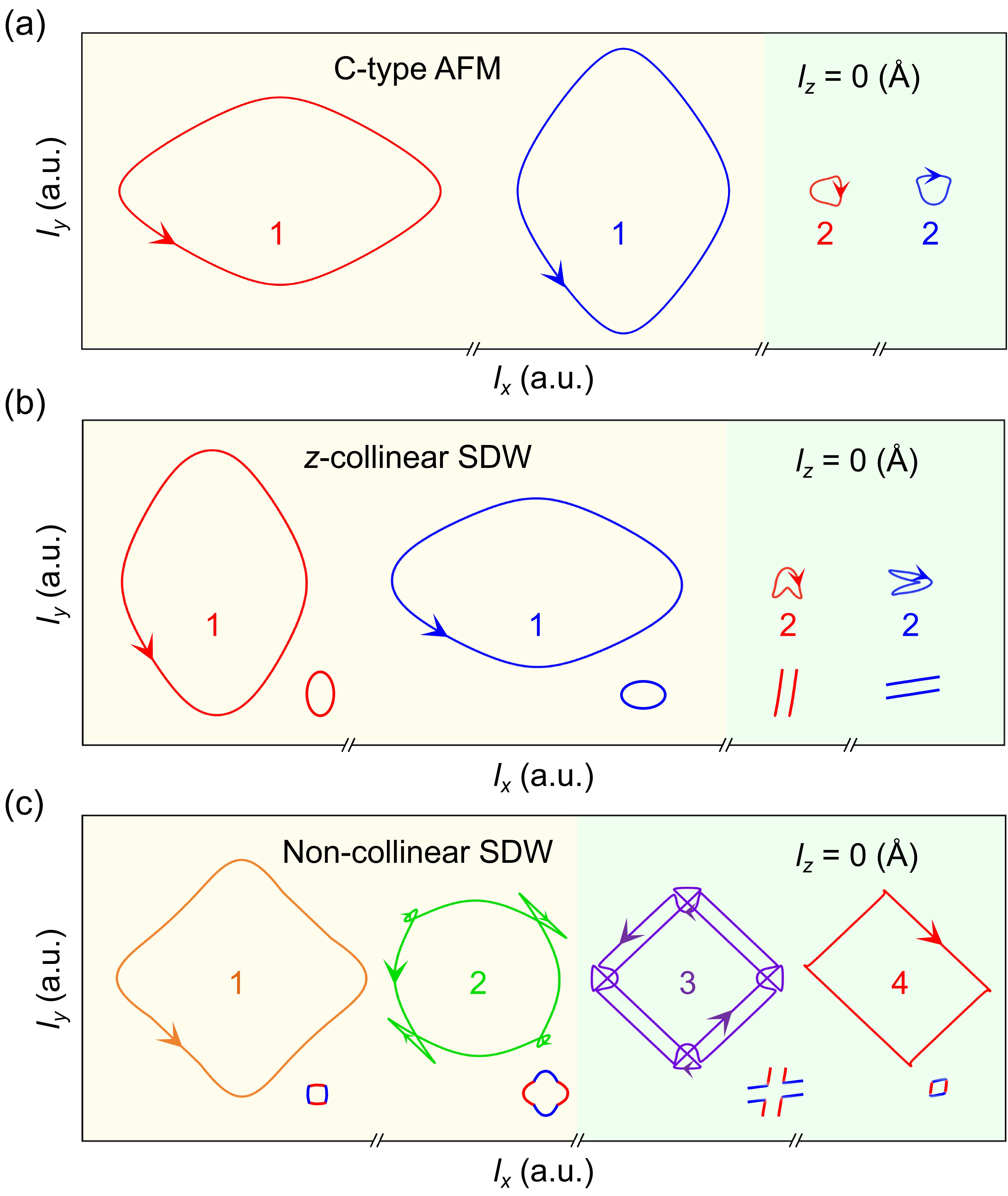}
    \caption{The \textbf{\textit{l}}-paths of carriers on Fermi surface with the magnetic field along \textit{z} direction of (a) the C-type AFM, (b) \textit{z}-collinear, and (c) non-collinear SDW KV\textsubscript{2}Se\textsubscript{2}O. The charge carriers moving clockwise and counterclockwise are electrons and holes, respectively. The colored numbers correspond to the band-indices in Figs. 2 and S2. The inserts show the corresponding Fermi surfaces.}
    \label{fig:3}
 \end{figure}

\textit{Carrier motion trajectory}—The scattering path length vector \textbf{\textit{l}} = \textbf{\textit{v}}(\textbf{\textit{k}})$\tau$ is introduced to maps the trajectory in momentum space to real space. Under out-of-plane magnetic field along \textit{z}-axis, the C-type AFM KV\textsubscript{2}Se\textsubscript{2}O exhibits spin-polarized carrier trajectories [Fig. 3(a)], in which holes in band 1 and electrons in band 2 follow counterclockwise and clockwise trajectories that resemble large diamond-like shapes and small shield-like patterns, respectively. The carrier trajectories in \textit{z}-collinear SDW state are similar to that in the C-type AFM state owing to the same Fermi surface topology [Fig. 3(b)]. For the non-collinear SDW KV\textsubscript{2}Se\textsubscript{2}O, the ``magnetic breakdown" phenomenon alters the carrier trajectories compare to that in the C-type AFM and \textit{z}-collinear SDW state [Fig. 3(c)]. Moreover, carriers in the near-circular profiles (band 1) and central parallelogram-shaped Fermi surface (band 4) exhibit hole and electron characteristics, respectively, while those in other Fermi surface (bands 2 and 3) exhibit alternating electron-like and hole-like characteristics during their propagation through real space. Therefore, KV\textsubscript{2}Se\textsubscript{2}O is expected to exhibit significant transformations in its magnetoelectric transport properties driven by phase transitions due to the change of carrier trajectories and concentrations. 

As shown in Fig. 1(c), the positive field dependence of $\rho_{yx}$ in both the C-type AFM and SDW states reveals hole-dominated transport in KV\textsubscript{2}Se\textsubscript{2}O. Moreover, the $\rho_{yx}$ of the non-collinear SDW KV\textsubscript{2}Se\textsubscript{2}O shows a nearly linear dependence on applied magnetic fields, reaching values an order of magnitude higher than that of the C-type AFM phase under strong field conditions. These obvious transport characteristics show remarkable agreement with the experimental results [Figs. 1(c) and 1(d)]\cite{bai2024absence}. The significant disparities in $\rho_{yx}$ can be illustrated based on the two-carrier model framework. For metal with both electron and hole carriers, the dependence of $\rho_{yx}$ on the magnetic field can be written as $\rho_{yx} = \frac{B \left( n_h \mu_h^2 - n_e \mu_e^2 \right) + \left( n_h - n_e \right) \mu_e^2 \mu_h^2 B^2}{e \left( n_e \mu_e + n_h \mu_h \right)^2 + \left( n_h - n_e \right)^2 \mu_e^2 \mu_h^2 B^2}$\cite{zhang2024complex}, in which \textit{$n_e $} and \textit{\textit{$n_h$} }representing the concentration of electron and hole charge carriers, respectively, and \textit{$\mu_e$} and \textit{$\mu_h $} denoting the carrier mobilities. In systems where one type of charge carrier dominates the transport, i.e $n_h \gg n_e$, the Hall resistivity can be illustrated as $\rho_{yx} \approx \frac{B}{e(n_h-n_e)}$ under strong magnetic field. In the non-collinear SDW KV\textsubscript{2}Se\textsubscript{2}O, ``magnetic breakdown" following the Fermi surface reconstruction induces carriers in petal-shaped Fermi surface (band 2) intermittent electron/hole behavior [Fig. 3(c)]. Compare to the \textit{z}-collinear state, this dynamic alternation increases the electron concentration on the combined elliptical Fermi surfaces [inserts in Fig. 3(b) and 3(c)], strengthening electron-hole compensation interactions and ultimately producing a significantly enhanced $\rho_{yx}$ [Fig. 1(c)]. The $\rho_{yx}$ of other magnetic symmetry breaking states also exhibits similar features (Fig. S3).

\textit{Anisotropy of magnetoresistance}—Spin canting induced ``magnetic breakdown" is further confirmed by comparing the anisotropic magnetoresistance (AMR) of the \textit{z}-collinear and \textit{z}-\textit{x}-non-collinear SDW KV\textsubscript{2}Se\textsubscript{2}O (Fig. S4). Here, $\theta$ is defined as the angle between the \textit{z}-axis and the direction of the applied magnetic field. The \textit{z}-collinear SDW state demonstrates a pronounced sensitivity to the field tilt direction [Fig. S4(a) and S4(b)]. Upon rotating the magnetic field from the \textit{z}-axis to the \textit{y}-axis, the magnetoresistance deviates sharply from a monotonic reduction and instead develops oscillatory modulations—a behavior that directly conflicts with experimental observations. Specifically, while the \textit{z}-collinear spin configuration maintains \textit{C}\textsubscript{2} rotational symmetry, magnetic field application along different crystallographic axis induces graded magnetic symmetry breaking—a mechanism manifested through filed-orientation-dependent MR. Conversely, for the non-collinear SDW phase, the positive MR exhibits a rapid suppression as the magnetic field tilting toward the crystallographic plane [Fig. S4(c) and S4(d)], and consistent with the experimental data neither the magnitude nor angular dependent evolution\cite{bai2024absence} [Fig. S4(e) and S4(f)]. These results conclusively reveal that the low-temperature transport behavior in KV\textsubscript{2}Se\textsubscript{2}O arises from ``magnetic breakdown"—a direct consequence of non-collinear-spin-driven magnetic symmetry breaking. 

\textit{Discussion}—This work resolves the experimental discrepancies\cite{bai2024absence,jiang2025metallic} of phase-dependent electronic and transport properties in KV\textsubscript{2}Se\textsubscript{2}O by establishing magnetic-group-symmetry dictated ``magnetic breakdown" as the governing mechanism. This anomalous transport mechanism is different from the weakened carrier scattering in SDW Mn\textsubscript{3}Si\cite{steckel2014spin}, metal-insulator transitions in CeOs\textsubscript{4}Sb\textsubscript{12}\cite{sugawara2005transport}, and alteration of carrier types induced by CDW transitions in TiSe\textsubscript{2}\cite{watson2019origin}. Here, the magnetic symmetry breaking along the $\Gamma$-M and X-$\Gamma$M/2 paths induced by spin canting is the most crucial factor for the spin-degenerate reduction and Fermi surface reconstruction. Notably, the transport sensitivity to 0.005 1/Å scale splitting, which may undetectable by ARPES, highlights the unique quantum metric response in altermagnets. 

The spin canting in the SDW state can be illustrated by the anisotropy of magnetic exchange interaction. This anisotropy originates from the periodic spin change along the SDW propagation vector, which breaking the \textit{C}\textsubscript{2\textit{z}} spin symmetry, thereby enhancing the directional dependence of the magnetic interactions. Consequently, the competition between the anisotropic exchange energy and the magnetocrystalline anisotropy induces the spins a slight deviation from the \textit{z}-axis\cite{akbari2011rkky,zapf2011varying}. The Fermi surface reconstruction induced ``magnetic breakdown" during the SDW phase transition presents a unique platform for temperature-controlled quantum manipulation, such as realizing nonlinear thermal transport through symmetry-selective gapping of electron-hole pockets and generating giant MR oscillations via reconfigurable Fermi surface nesting. Such distinct transport signatures show the potential to enhance the sensitivity and selectivity of magnetic field detection in advanced sensor applications. The causal relationships among ``magnetic breakdown", Fermi surface topology, and emergent transport anomalies remain to be elucidated, which would further emphasize the critical role of Fermi surface topology in governing quantum properties. 

\textit{Conclusion}—To resolve the paradox between conserved Fermi surface by ARPES and distinct transport behaviors across the SDW transition in d-wave altermagnetic KV\textsubscript{2}Se\textsubscript{2}O, we systematically deciphers the interplay between magnetic phase transitions, spin alignment, Fermi surface topology, and transport properties in the multi-state altermagnetic KV\textsubscript{2}Se\textsubscript{2}O. By linking spin symmetry evolution with momentum-locked spin-polarized bands, we demonstrate that the spin canting triggers symmetry-constrained Fermi surface reconstruction and spin flip. The resulting ``magnetic breakdown" phenomenon strengthens electron-hole compensation and governs unique transport characteristics. The concordance between theoretically calculated and experimentally measured values of both the $\rho_{yx}$ and AMR serves to unequivocally validate the SDW driven ``magnetic breakdown" in KV\textsubscript{2}Se\textsubscript{2}O. Our work proposes a framework for understanding phase dependent transport characteristic in altermagents and provides a novel perspective for designing symmetry-tunable spintronic devices based on altermagnetic phase transitions. 

\textit{Data availability statement}—The data required to reproduce the DFT and magnetotransport calculations is available in the Electronic Laboratory for Materials Science, hosted on the website of the Condensed Matter Physics Data Center, Chinese Academy of Sciences: \url{https://in.iphy.ac.cn/eln/link.html#/115/W2j7}.

\textit{Acknowledgments}—This work was supported by the National Natural Science Foundation of China (Grants No. 12274436 and No. 11921004), the National Key R\&D Program of China (Grants No. 2023YFA1607400 and No. 2022YFA1403800), the Science Center of the National Natural Science Foundation of China (Grant No. 12188101), the Central Guiding Local Science and Technology Development Fund Projects (Grant No. 2111400031). H.W. acknowledges support from the New Cornerstone Science Foundation through the XPLORER PRIZE.

%

\end{document}